\documentclass[aps,prb,reprint,superscriptaddress]{revtex4-2}   
\usepackage{physics,amssymb,graphicx,bm,microtype,siunitx,hyperref}
\hypersetup{hidelinks}
\newcommand{\mr}[1]{\mathrm{#1}}
\newcommand{\eq}[1]{\text{$#1$}}

\begin{document}
\title{Light-induced rectified orbital magnetization in electron-hole bilayers}

\author{Erlend Syljuåsen}
\affiliation{Center for Quantum Spintronics, Department of Physics, Norwegian University of Science and Technology, NO-7491 Trondheim, Norway}
\author{Gabriel Cardoso}
\affiliation{Nordita, Stockholm University, and KTH Royal Institute of Technology, 
Hannes Alfv\'ens v\"ag 12, SE-106 91 Stockholm, Sweden}
\author{Esra Ilke Albar}
\affiliation{Fritz-Haber-Institute of the Max-Planck-Society, 14195 Berlin, Germany}
\author{\\Patrick J. Wong}
\affiliation{Nordita, Stockholm University, and KTH Royal Institute of Technology,
Hannes Alfv\'ens v\"ag 12, SE-106 91 Stockholm, Sweden}
\affiliation{Department of Physics, University of Connecticut, Storrs, Connecticut 06269, USA}
\author{Angel Rubio}
\affiliation{Max Planck Institute for Structure and Dynamics of Matter, Center for Free-Electron Laser Science, Luruper Chaussee 149, 22761 Hamburg, Germany}
\affiliation{Initiative for Computational Catalysis (ICC) and Center for Computational Quantum Physics, Flatiron Institute, 162 5th Avenue, New York, NY 10010.}
\author{Alexander V. Balatsky}
\affiliation{Nordita, Stockholm University, and KTH Royal Institute of Technology,
Hannes Alfv\'ens v\"ag 12, SE-106 91 Stockholm, Sweden}
\affiliation{Department of Physics, University of Connecticut, Storrs, Connecticut 06269, USA}

\begin{abstract}
Circularly polarized light can rectify orbital motion into a static magnetization through the inverse Faraday effect (IFE), but in electron-hole bilayers the electron and hole contributions cancel exactly when their properties are equivalent. We show that electron-hole bilayers in transition-metal dichalcogenide platforms avoid this cancellation through effective-mass asymmetry alone, and that the surviving orbital IFE is sensitive to interlayer coupling. In the weak-coupling regime, a two-component Drude description yields an induced magnetization of order one Bohr magneton per carrier for representative terahertz driving. In the strong-coupling regime, where the carriers bind into interlayer excitons, we treat the relative motion using a hydrogenic model with a Rytova-Keldysh interaction and obtain a reduced but finite response. We give closed-form expressions that allow estimates across a broad parameter range and identify where the response is largest.
\end{abstract}
\maketitle

\section{Introduction}
Coupling matter to tailored electromagnetic fields enables control and detection of material properties. A direct realization occurs when properties of the drive, such as the angular momentum of light, are imprinted onto matter \cite{aeppli_2025}. Rectification effects are of particular interest, as nonlinearities can generate a static response from an oscillating drive. Relevant for our discussion is the example of inverse Faraday effect (IFE), in which circularly polarized light generates a static magnetization in a material \cite{pitaevskii_1960, pershan_1963, ziel_1965, pershan_1966}. 

The IFE-induced magnetization can have both spin and orbital components \cite{tazuke_2025}. We focus on the orbital component, which can be understood as rectified light-driven carrier motion: a rotating electric field generates static vortical currents inducing an orbital magnetic moment. Controlling this orbital response enables optical manipulation of magnetization without static magnetic fields. Sensitivity of orbital motion to the electronic structure also makes it a probe of carrier dynamics \cite{hareau_2025}. Previous studies of the orbital IFE span plasma physics \cite{deschamps_1970, belkov_1979, kono_1981, karpman_1982, horovitz_1997, najmudin_2001}, normal metals \cite{hertel_2006, taguchi_2011, berritta_2016, tanaka_2020, karakhanyan_2022, sharma_2024}, topological systems \cite{misawa_2011, tokman_2020, liang_2021, zhang_2023, cardoso_2026}, and superconductors \cite{mironov_2021, majedi_2021, putilov_2023, dzero_2024}. These studies have largely focused on systems whose mobile carriers carry a net charge.

In this article, we consider an electron-hole bilayer and show that it can support an orbital IFE. The system consists of electron-doped and hole-doped layers separated by an insulating spacer whose thickness is much smaller than the wavelength of the applied light. The spacer suppresses direct interlayer tunneling while retaining the Coulomb attraction between the layers, and electrostatic gates allow the carrier densities to be controlled independently. Such electron-hole bilayer structures have been proposed as platforms for exciton condensation \cite{fang_2014, ming_2018, forg_2019, jiang_2021, ma_2021, wu_2015}. We consider the specific case of transition-metal dichalcogenide (TMD) monolayers, although our mechanism for the orbital IFE applies more generally to spatially separated two-dimensional electron and hole liquids. 

\begin{figure}[htb]
\includegraphics[width=0.9\linewidth]{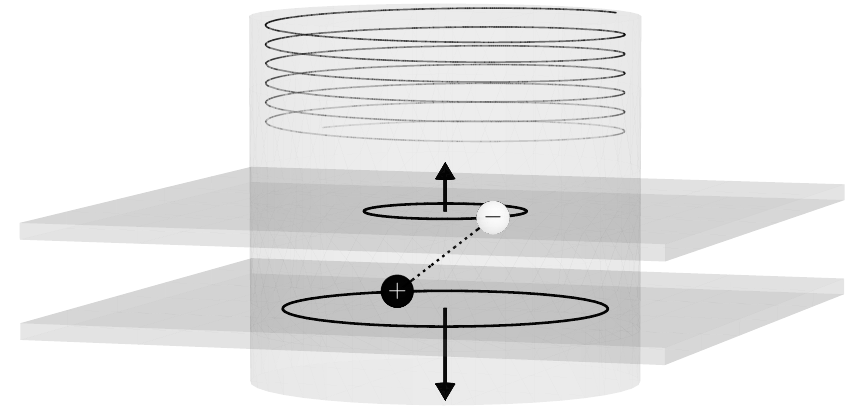}
\caption{Schematic of the orbital IFE in an electron-hole bilayer. Electrons in the top layer (open sphere) and holes in the bottom layer (filled sphere) are driven by circularly polarized light (shaded cylindrical region). Electron-hole asymmetry, such as unequal effective masses, prevents exact cancellation of their contributions to the rectified orbital magnetic moment.}
\label{fig:bilayer}
\end{figure}

The IFE requires active ``stirring'' of carriers by light. For a compensated bilayer with no net charge and electron-hole spacing much smaller than the wavelength of the light, the light effectively probes a charge-neutral system. Still we find a finite IFE for generic electron-hole carrier asymmetry. This asymmetry can arise from differences in carrier density or relaxation time, while in compensated TMD heterostructures it occurs naturally due to unequal electron and hole effective masses \cite{mostaani_2017, dai_2024, fu_2026}. TMDs are often discussed in terms of valley and spin degrees of freedom. Here, we instead focus on the orbital contribution to a generic IFE, Fig.~\ref{fig:bilayer}, and describe the carrier dynamics using a continuum effective-mass model. Related orbital IFE mechanisms have recently been discussed in the context of dynamical multiferroicity \cite{basini_2024} and Rydberg systems \cite{wong_2025}. 

In the weak interlayer coupling regime, we use a two-component Drude model with interlayer Coulomb drag, and find an IFE-induced magnetization of order one Bohr magneton per carrier for terahertz driving with field amplitude \eq{E_0 = \SI{e5}{\volt\per\meter}}. This estimate uses an effective-mass ratio \eq{m_\mr{e} / m_\mr{h} = 2} representative of a \eq{\mr{WSe_2/MoSe_2}} bilayer \cite{yeh_2015, mostaani_2017}. In the opposite limit of strong interlayer coupling, the magnetization is substantially suppressed as the electrons and holes become locked by the Coulomb attraction. In this limit, quantum effects of interlayer excitons become important, which we describe using a hydrogenic Hamiltonian with a Rytova-Keldysh screened interaction. We find a smaller but finite IFE using numerical time evolution. We then account for this result using time-dependent perturbation theory and derive a compact off-resonant expression that enables estimates over a range of parameters. This paper is organized as follows: in section~\ref{sec:two_component_drude_model} we study the IFE using the two-component Drude model, and in section~\ref{sec:quantum-exciton} we use the quantum model of interlayer excitons. We give a summary of our findings in section~\ref{sec:summary}.

\section{Two-component Drude model} \label{sec:two_component_drude_model}
We begin in the weak interlayer coupling regime and use a two-component description characterized by densities \eq{n_\alpha(\bm{r}, t)} and velocities \eq{\bm{v}_\alpha(\bm{r}, t)}, where \eq{\alpha = \mr{e}} refers to electrons in layer 1 and \eq{\alpha = \mr{h}} to holes in layer 2. These fields satisfy the continuity equation
\begin{equation} \label{eq:continuity}
	\dot{n}_\alpha + \div (n_\alpha \bm{v}_\alpha) = 0.
\end{equation}
The charge current density of component \eq{\alpha} is \eq{\bm{J}_\alpha = q_\alpha n_\alpha \bm{v}_\alpha}, with charges \eq{q_\mr{e}=-e} and \eq{q_\mr{h}=e}, where \eq{e > 0} is the elementary charge. Supplementing this description with a momentum-balance equation relating the velocities to the applied electric field, the IFE appears as a rectified solenoidal contribution to the charge current. The associated IFE-induced magnetization can then be identified from this current contribution \cite{karpman_1982, hertel_2006, cardoso_2026}. Before computing the magnetization, we present and analyze the momentum-balance equations relevant for electron-hole bilayers.

\subsection{Equations of motion}
The Drude model of transport in bilayer systems incorporates interlayer Coulomb interactions through a drag term in the equations of motion \cite{pogrebinskii_1977, narozhny_2016}. In the spatially uniform limit, the coupled equations read
\begin{subequations} \label{eq:drude_eom}
\begin{align}
	m_\mr{e} \dot{\bm{v}}_\mr{e}  = q_\mr{e} \bm{E}& - \frac{\gamma}{n_\mr{e}}(\bm{v}_\mr{e} - \bm{v}_\mr{h}) - \frac{m_\mr{e}}{\tau_\mr{e}} \bm{v}_\mr{e},
	\\
	m_\mr{h} \dot{\bm{v}}_\mr{h} = q_\mr{h} \bm{E}& - \frac{\gamma}{n_\mr{h}}(\bm{v}_\mr{h} - \bm{v}_\mr{e}) - \frac{m_\mr{h}}{\tau_\mr{h}} \bm{v}_\mr{h},
\end{align}
\end{subequations}
where \eq{m_\alpha} is the effective mass and \eq{\tau_\alpha} the intralayer scattering time for component \eq{\alpha}. The parameter \eq{\gamma > 0} sets the strength of the interlayer Coulomb drag, and the factors \eq{1/n_\alpha} make the force densities equal and opposite, such that interlayer drag alone conserves total momentum. The interlayer drag term transfers momentum between the electron and hole fluids and acts as an effective frictional coupling that relaxes the relative velocity. We consider an externally applied monochromatic electric field with angular frequency \eq{\omega}, written in complex notation as \eq{\bm{E}(t)=\Re(\bm{E}^{\omega} e^{i\omega t})}. Using the trial solution \eq{\bm{v}_\alpha(t)=\Re(\bm{v}^\omega_\alpha e^{i\omega t})} in Eq.~\eqref{eq:drude_eom} gives
\begin{align} \label{eq:drude_velocities}
	\bm{v}^\omega_\mr{e} &= \frac{\left(L_\mr{h} q_\mr{e} + \gamma_\mr{e} q_\mr{h} \right) \bm{E}^\omega }{L_\mr{e} L_\mr{h} - \gamma_\mr{e} \gamma_\mr{h}}, 
	&
	\bm{v}^\omega_\mr{h} &= \frac{\left(L_\mr{e} q_\mr{h} + \gamma_\mr{h} q_\mr{e} \right) \bm{E}^\omega}{L_\mr{e} L_\mr{h} - \gamma_\mr{e} \gamma_\mr{h}},
\end{align}
where
\begin{align} \label{eq:gamma_L_definitions}
	\gamma_\alpha &= \frac{\gamma}{n_\alpha},
	&
	L_\alpha &= i\omega m_\alpha + \gamma_\alpha + \frac{m_\alpha}{\tau_\alpha}.
\end{align}
To clarify the behavior analytically, we consider two limiting cases: vanishing drag \eq{(\gamma = 0)} and strong drag \eq{(\gamma \to \infty)}. We emphasize that the latter limit is used only to illustrate the locking of electron and hole motion, whereas strong interlayer Coulomb correlations and exciton formation are treated separately later.

\begin{figure}[htb]
\includegraphics[width=0.94\linewidth]{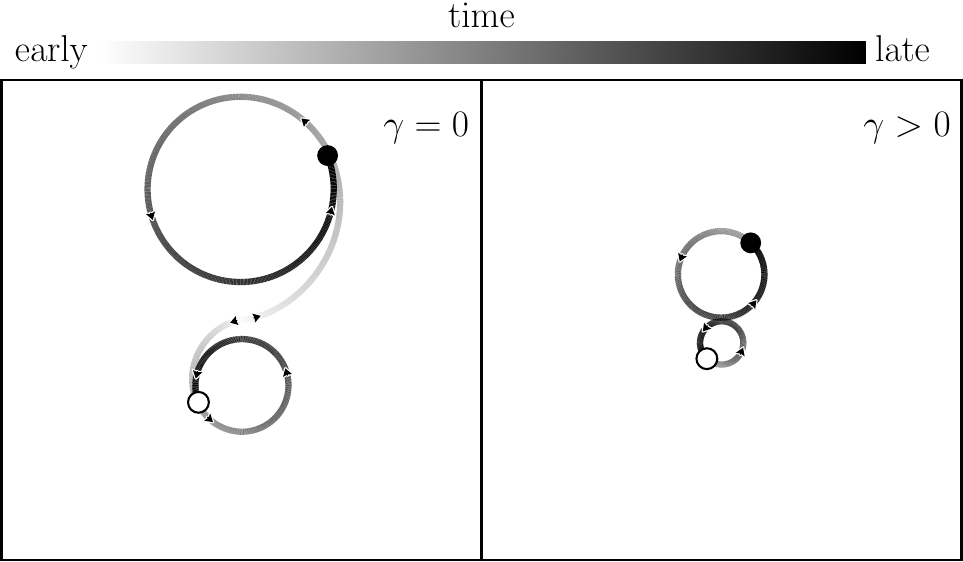}
\caption{Schematic of electron (open circle) and hole (filled circle) trajectories driven by circularly polarized light for vanishing (\eq{\gamma = 0}) and finite (\eq{\gamma > 0}) interlayer drag. The curves show a two-dimensional projection of the in-plane motion and the color gradient indicates time evolution with both carriers starting at the origin.}
\label{fig:trajectories}
\end{figure}
\vspace{-0.1cm}
In the absence of interlayer drag, the above solutions decouple and reduce to the standard Drude response,
\begin{align} \label{eq:vanishing_drag}
	\bm{v}^\omega_{\mr{e}} &\underset{\gamma = 0}{=} \frac{q_\mr{e} \tau_\mr{e} \bm{E}^\omega}{m_\mr{e}(1 + i \omega \tau_\mr{e})}, 
	&
	\bm{v}^\omega_{\mr{h}} &\underset{\gamma = 0}{=} \frac{q_\mr{h} \tau_\mr{h} \bm{E}^\omega}{m_\mr{h}(1 + i\omega \tau_\mr{h})}.
\end{align}
For circularly polarized driving, Eq.~\eqref{eq:vanishing_drag} shows that electrons and holes rotate in the same direction set by the handedness of the drive, but their motions are phase-shifted by \eq{\pi} because \eq{q_\mr{e} = -q_\mr{h}}. The relevant parameter controlling the difference between the electron and hole responses in Eq.~\eqref{eq:vanishing_drag} is the ratio \eq{\tau_\alpha / m_\alpha}, which in TMD bilayers can differ substantially due to unequal effective masses. In the strong-drag limit, the two fluids are locked and move with a common velocity,
\begin{align} \label{eq:vdipole}
	\bm{v}^\omega_{\mr{e}} = \bm{v}^\omega_{\mr{h}} \! \underset{\gamma \to \infty}{=} \frac{\tau_\mr{e} \tau_\mr{h}(n_\mr{e} q_\mr{e} + n_\mr{h} q_\mr{h} ) \bm{E}^\omega}{n_\mr{e} m_\mr{e} \tau_\mr{h}(1 + i\omega \tau_\mr{e}) + n_\mr{h} m_\mr{h} \tau_\mr{e}(1 + i\omega \tau_\mr{h})}.
\end{align}
Strong drag thus causes co-motion, with the velocity set by the density imbalance. As a result, we expect a suppressed net response for charge-balanced bilayers in the strong-drag limit. 
Figure~\ref{fig:trajectories} shows numerical solutions of Eq.~\eqref{eq:drude_eom} for circularly polarized driving, illustrating the effect of interlayer drag. Finite intralayer momentum relaxation leads to circular orbits in both cases, with the electron and hole rotating in the same direction. We used \eq{m_\mr{e} > m_\mr{h}} and \eq{\tau_\mr{e} = \tau_\mr{h}} in the figure, giving a slightly larger hole orbital radius. Increasing interlayer drag couples the two motions and reduces the radii of rotation. However, increasing interlayer Coulomb coupling eventually takes the system beyond the regime of weakly correlated carriers. The Drude model cannot capture the resulting bound-state formation and the corresponding discrete energy levels, and a quantum treatment of the electron-hole motion is therefore required to describe the interlayer excitonic dynamics. We will later analyze this regime in detail in Sec.~\ref{sec:quantum-exciton}.

\subsection{Coulomb drag resistivity}
The interlayer drag coefficient \eq{\gamma} can be related to the measured drag resistivity. In a standard Coulomb drag experiment \cite{narozhny_2016}, a current is driven through one (active) layer, and interlayer momentum transfer drives carriers in the other (passive) layer. Because the passive layer is held under open-circuit conditions, charge accumulates until an internal electric field builds up that cancels the drag force in the steady state. The resulting electrostatic potential difference defines the drag voltage in the passive layer \cite{narozhny_2016}. 

We take layer 2 (holes) as the passive layer and define the drag resistivity as 
\begin{equation}
\rho_\mr{D} = \frac{E_\mr{h}}{J_\mr{e}} \qq{with} J_\mr{h} = 0.
\end{equation}
Here, \eq{E_\mr{h}} is the longitudinal electric field induced in the passive layer, \eq{J_\mr{e}} is the applied longitudinal charge current density in the active layer, and \eq{J_\mr{h} = 0} enforces open-circuit conditions in the passive layer. Comparing with Eqs.~\eqref{eq:drude_eom}, we replace the driving terms with static longitudinal fields, \eq{\bm{E}_\alpha \to \hat{\bm{x}} E_\alpha}, impose \eq{J_\mr{h} = 0} and the steady-state conditions \eq{\dot{v}_\mr{e} = \dot{v}_\mr{h} = 0}. The passive-layer equation gives \eq{E_\mr{h} = -\gamma v_\mr{e} / (q_\mr{h} n_\mr{h})}, and using \eq{J_\mr{e} = q_\mr{e} n_\mr{e} v_\mr{e}}, we obtain
\begin{equation}
\rho_\mr{D} = -\frac{\gamma}{q_\mr{e} q_\mr{h} n_\mr{e} n_\mr{h}}.
\end{equation}
This relation is useful because measurements of \eq{\rho_{\mr{D}}} have been used to distinguish weakly coupled Fermi-liquid behavior from regimes with strong electron-hole correlations and exciton formation \cite{narozhny_2016, croxall_2008, seamons_2009}. In the following, we therefore use \eq{\rho_\mr{D}} as a scale to track the crossover towards more strongly correlated interlayer dynamics.

\subsection{Rectified magnetization}
We now compute the rectified orbital magnetization induced by the IFE. The derivation of the rectified magnetization in terms of the first-harmonic velocity components is presented in Appendix~\ref{appendix:ife}, following Ref.~\cite{karpman_1982} and the later formulations in Refs.~\cite{hertel_2006, cardoso_2026}. The net rectified out-of-plane magnetization, obtained as the combined contribution of both layers, is
\begin{equation} \label{eq:result}
    M = \sum_\alpha  \frac{q_\alpha n_\alpha}{4\omega} \Im(\bm{v}^\omega_{\alpha} \cross \bm{v}^{\omega *}_{\alpha})_z = \chi  \Im(\bm{E}^\omega \cross \bm{E}^{\omega *})_z,
\end{equation}
where the second equality follows by substituting the velocities in Eq.~\eqref{eq:drude_velocities}, and the IFE response function is
\begin{equation}
	\chi = \frac{q_\mr{e} n_\mr{e} \abs{L_\mr{h} q_\mr{e} + \gamma_\mr{e} q_\mr{h}}^2 + q_\mr{h} n_\mr{h} \abs{L_\mr{e} q_\mr{h} + \gamma_\mr{h} q_\mr{e}}^2}{4\omega \abs{L_\mr{e} L_\mr{h} - \gamma_\mr{e} \gamma_\mr{h}}^2}.
\end{equation}
To make the helicity dependence of the applied field explicit in Eq.~\eqref{eq:result}, we decompose the complex field amplitudes into circular components \eq{E^\omega_\pm = (E^\omega_x \pm i E^\omega_y) / \sqrt{2}} and introduce the helicity factor
\begin{equation} \label{eq:helicity}
	\mathcal{P} = \frac{\abs{E^\omega_+}^2 - \abs{E^\omega_-}^2}{\abs{E^\omega_+}^2 + \abs{E^\omega_-}^2} \in \left[-1, 1\right].
\end{equation}
Purely circular polarization corresponds to \eq{\mathcal{P} = \pm 1}, while linear polarization corresponds to \eq{\mathcal{P} = 0}. We take \eq{\mathcal{P} = 1} to denote left-handed circular polarization (LHCP) and \eq{\mathcal{P} = -1} to denote right-handed circular polarization (RHCP), for light incident from above onto the sample with the handedness defined as from the point of view of the source. The helicity dependence in Eq.~\eqref{eq:result} follows from
\begin{equation} \label{eq:ife_depend}
    \Im(\bm{E}^\omega \cross \bm{E}^{\omega *})_z = \abs{E^\omega}^2 \mathcal{P},
\end{equation}
where \eq{\abs{E^\omega}^2 = \abs{E^\omega_+}^2 + \abs{E^\omega_-}^2}, which in this convention is related to the real-time amplitude by \eq{E_0^2 = \abs{E^\omega}^2/2}. Equation~\eqref{eq:result} together with Eq.~\eqref{eq:ife_depend} thus shows that a nonzero magnetization requires a field with finite helicity, and that its sign reverses when the helicity is switched.

We next analyze the IFE response function in the two limiting cases of vanishing and strong interlayer drag,
\begin{subequations} \label{eq:ife_response_limits}
\begin{align}
	&\chi \underset{\gamma = 0}{=} \frac{n_\mr{e} q_\mr{e}^3 \tau_\mr{e}^2}{4\omega m_\mr{e}^2 \left(1 + \omega^2 \tau_\mr{e}^2 \right)} + \frac{n_\mr{h} q_\mr{h}^3 \tau_\mr{h}^2}{4\omega m_\mr{h}^2 \left(1 + \omega^2 \tau_\mr{h}^2 \right)},
	\\
	&\chi \underset{\gamma \to \infty}{=} \frac{\left(n_\mr{e} q_\mr{e} + n_\mr{h} q_\mr{h} \right)^3}{4\omega[\left(m_\mr{e} n_\mr{e} / \tau_\mr{e} + m_\mr{h} n_\mr{h} / \tau_\mr{h} \right)^2 + (m_\mr{e} n_\mr{e} + m_\mr{h} n_\mr{h})^2 \omega^2]}.
\end{align}
\end{subequations}
In the absence of interlayer drag \eq{(\gamma = 0)}, the magnetization reduces to the difference between the independent electron and hole responses. Since \eq{q_\mr{e} = -q_\mr{h}}, the two contributions cancel when the electron and hole responses are identical. A finite magnetization therefore requires asymmetry in carrier densities, effective masses, or scattering times. In the strong-drag limit \eq{(\gamma \to \infty)}, the electron and hole become locked, and a nonzero magnetization arises only for unequal densities.

\begin{figure}[htb]
    \includegraphics[width=0.99\linewidth]{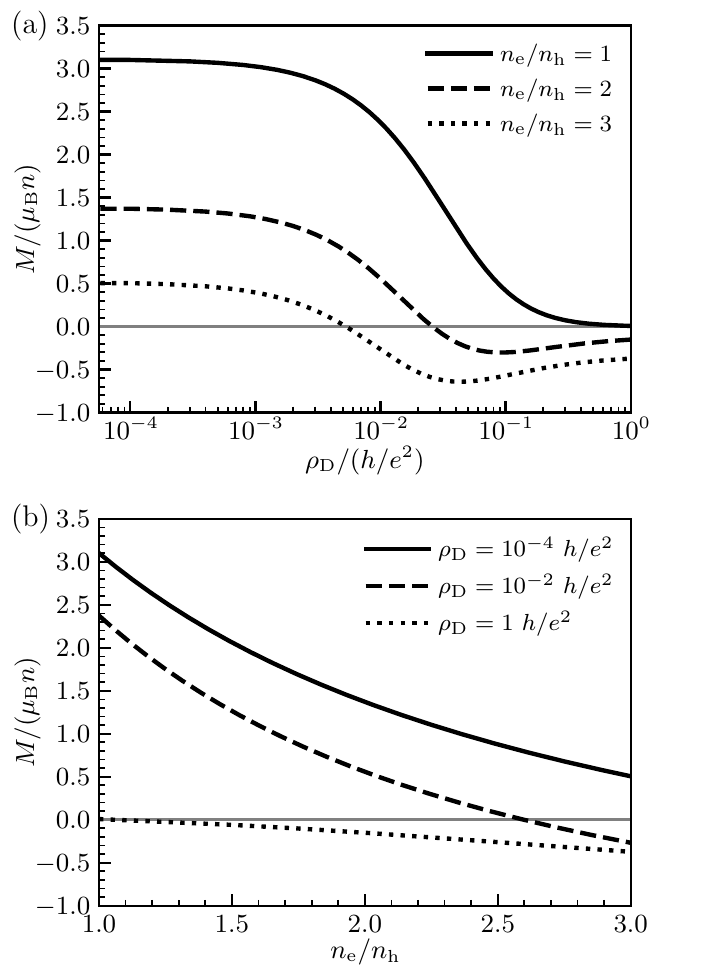}
    \caption{Estimated IFE-induced magnetization in a TMD electron-hole bilayer as a function of (a) drag resistivity and (b) density ratio. The effective masses are \eq{m_\mathrm{e} = 0.8 m_0} and \eq{m_\mathrm{h} = 0.4m_0}, where \eq{m_0} is the free electron mass. The total density is \eq{n = n_e + n_h}, with \eq{n_\mr{h} = \SI{e12}{\per\square\centi\meter}}. We take equal intralayer scattering times, \eq{\tau_\mr{e} = \tau_\mr{h} = \SI{e-12}{\second}}. The applied electric field is LHCP with angular frequency \eq{\omega = \SI{e12}{\per\second}} and amplitude \eq{E_0 = \SI{e5}{\volt\per\meter}}.}
    \label{fig:ife_classical}
\end{figure}

Figure~\ref{fig:ife_classical}~(a) shows the IFE-induced magnetization as a function of the Coulomb drag resistivity \eq{\rho_\mr{D}} for parameters representative of a TMD electron-hole bilayer. For equal carrier densities (solid line), the imbalance in effective masses produces a magnetization on the order of one Bohr magneton per carrier in the weak-drag regime. The magnetization starts to decrease near \eq{\rho_\mr{D} \approx 10^{-2} h / e^2}. This scale corresponds to the crossover where the interlayer drag term \eq{\gamma / n_\alpha} becomes comparable to the intralayer momentum relaxation term \eq{m_\alpha / \tau_\alpha} in Eq.~\eqref{eq:drude_eom}, that is, where interlayer drag competes with intralayer dissipation. The magnetization vanishes in the strong-drag regime for equal carrier densities, where interlayer momentum transfer dominates over intralayer relaxation. For the parameters shown, this occurs when \eq{\rho_\mr{D}} becomes a sizable fraction of the two-dimensional resistance quantum \eq{h / e^2}.

So far, we have focused on mass asymmetry alone, but density imbalance provides an additional source of electron-hole asymmetry and can also contribute to a finite IFE. The dashed curves in Fig.~\ref{fig:ife_classical}~(a) and Fig.~\ref{fig:ife_classical}~(b) show that density imbalance permits finite IFE-induced magnetization even in the strong-drag limit, consistent with the velocity analysis in Eq.~\eqref{eq:vdipole}. In addition, the imbalance can lead to a sign reversal of the magnetization as \eq{\rho_\mr{D}} is increased. This reversal reflects the competition between electron and hole contributions in the weak-drag and strong-drag regimes, and can therefore provide a signature of the crossover from independent Drude motion to drag-locked dynamics. We now derive the condition for this sign reversal. For equal intralayer scattering times, Eq.~\eqref{eq:ife_response_limits} gives \eq{\text{sgn}(\chi|_{\gamma = 0}) = \text{sgn}(n_\mr{h} / m^2_\mr{h} - n_\mr{e} / m^2_\mr{e})} and \eq{\text{sgn}(\chi|_{\gamma \to \infty}) = \text{sgn}(n_\mr{h} - n_\mr{e})}. Defining the proportionality constants \eq{c_m = m_\mr{e} / m_\mr{h}} and \eq{c_n = n_\mr{e} / n_\mr{h}} satisfying \eq{c_n > 0} and \eq{c_m > 0}, we find a sign reversal between the weak-drag and strong-drag regimes when either \eq{c^2_m < c_n < 1} or \eq{c^2_m > c_n > 1}. 

\section{Quantum exciton model} \label{sec:quantum-exciton}
The two-component Drude model considered above treats the two layers as independent semiclassical carrier fluids coupled by frictional drag. This description is expected to be valid in the weak interlayer coupling regime, where the drag resistivity is small. Extending this model towards the stronger drag regime, we found a suppression of the IFE and a vanishing induced magnetization for charge-balanced bilayers. This behavior can be understood from a dipolar picture: in the charge-balanced strong-drag limit, the electron and hole become locked together so that their in-plane separation vanishes. As shown in Appendix~\ref{app:classical_dipole_limit}, a finite orbital magnetic moment requires a finite in-plane separation, and the IFE is therefore suppressed in this limit.

However, this argument does not capture quantum effects associated with the formation of bound electron-hole excitons. In this section, we investigate the excitonic regime and find a reduced but finite orbital IFE. Electron-hole asymmetry remains necessary in this regime, where we here also consider unequal electron and hole effective masses. We now explain the origin of this effect and estimate its magnitude for representative TMD bilayers.

\subsection{Interlayer exciton Hamiltonian}
The many-body excitonic problem can be approximated by an effective Schr{\"o}dinger equation for a single electron-hole pair with a hydrogenic Hamiltonian \cite{wannier_1937}, assuming the dilute regime where electrons and holes form well-defined spatially extended bound states. This continuum Wannier-Mott description is appropriate when the exciton radius is large compared with the lattice constant and exciton-exciton interactions are negligible \cite{kamban_2020}. It has been widely used to model the optical properties of excitons \cite{quintela_2022}, making it a natural starting point for describing the IFE in the regime of strong interlayer correlations.

We consider a single electron-hole pair described by the Hamiltonian
\begin{equation}
    H(t) = \frac{\bm{p}^2_\mr{e}}{2m_\mr{e}} + \frac{\bm{p}^2_\mr{h}}{2m_\mr{h}} + V(|\bm{r}_\mr{e} - \bm{r}_\mr{h}|) +e(\bm{r}_\mr{e}  - \bm{r}_\mr{h}) \vdot \bm{E}(t),
\end{equation}
where \eq{\bm{r}_\alpha} and \eq{\bm{p}_\alpha} are the position and momentum operators of the electron \eq{(\alpha = \mr{e})} and hole \eq{(\alpha = \mr{h})}. The operators \eq{\bm{r}_\alpha} and \eq{\bm{p}_\alpha} act within the two-dimensional layers, while the layer spacing \eq{d} enters as a fixed parameter in the potential. We assume the applied electric field \eq{\bm{E}(t)} is spatially uniform on the scale of the exciton and write the light-matter interaction in the electric-dipole approximation. The Coulomb interaction in two-dimensional semiconductors is well described by the Rytova-Keldysh potential \cite{rytova_1967, keldysh_1979, cudazzo_2010, cudazzo_2011}, since the bare \eq{1/r} interaction is modified by dielectric screening in atomically thin layers. It has been widely used to model exciton states in monolayer TMDs \cite{berkelbach_2013, chernikov_2014, tuan_2018} and has recently also been extended to bilayer TMDs \cite{kamban_2020}. Following Ref.~\cite{kamban_2020}, we write the potential as
\begin{equation}
	V(r) = -\frac{e^2}{4 \pi \varepsilon_0} \frac{\pi}{2 r_0} \left[ H_0\left(\frac{\rho}{r_0}\right) - Y_0\left(\frac{\rho}{r_0} \right) \right],
\end{equation}
where \eq{H_0} is the zeroth-order Struve function, \eq{Y_0} is the zeroth-order Neumann function (Bessel-\eq{Y}), \eq{r_0} denotes the combined screening length of the two layers, and \eq{\varepsilon_0} is the vacuum permittivity. The effective radial separation
\begin{equation}
	\rho = \kappa \sqrt{r^2 + d^2},
\end{equation}
includes the fixed interlayer distance \eq{d} and the dielectric environment through \eq{\kappa = (\varepsilon_+ + \varepsilon_-) / 2}, where \eq{\varepsilon_+} (\eq{\varepsilon_-}) is the relative dielectric constant above (below) the bilayer. Reference~\cite{cudazzo_2011} shows that the asymptotic expansions of the zeroth-order Struve and Neumann functions give
\begin{subequations}
\begin{align}
	V(r) \underset{\rho \gg r_0}{\approx}& -\frac{e^2}{4\pi \varepsilon_0} \frac{1}{\rho},
	\\
	V(r) \underset{\rho \ll r_0}{\approx}& -\frac{e^2}{4\pi \varepsilon_0} \frac{1}{r_0} \left[\ln(\frac{2r_0}{\rho}) - \gamma_\mr{EM} \right],
\end{align}
\end{subequations}
where \eq{\gamma_\mr{EM}} is the Euler-Mascheroni constant. The attractive interaction therefore approaches a Coulomb-like form \eq{(\sim 1/\rho)} at large distances and becomes logarithmic at short distances.

We next introduce the relative and center-of-mass (COM) coordinates
\begin{align}
	\bm{r} &= \bm{r}_\mr{e} - \bm{r}_\mr{h},
	\\
	\bm{R} &= \frac{1}{m_\mr{tot}} \left( m_\mr{e} \bm{r}_\mr{e} + m_\mr{h} \bm{r}_\mr{h} \right),
\end{align}
where \eq{m_\mr{tot} = m_\mr{e} + m_\mr{h}}. In these variables, the Hamiltonian becomes
\begin{equation} \label{eq:relative_H}
    H(t) = \frac{\bm{P}^2}{2m_\mr{tot}} + \frac{\bm{p}^2}{2\mu} + V(r) + e \bm{r} \vdot \bm{E}(t),
\end{equation}
where \eq{1/\mu = 1/m_\mr{e} + 1/m_\mr{h}} is the inverse reduced mass, and the operators \eq{\bm{p}} and \eq{\bm{P}} are the canonical conjugate momenta to \eq{\bm{r}} and \eq{\bm{R}}, respectively. Equation~\eqref{eq:relative_H} shows that the COM part decouples from the internal exciton dynamics. We therefore focus on the relative-motion Hamiltonian, which determines the exciton spectrum and the effect of the coupling to the electric field.

\subsection{Binding energy and excited states} 
Before turning to the IFE-induced magnetization, we determine the equilibrium interlayer exciton spectrum that sets the relevant energy scale. We obtain the eigenenergies and eigenstates by numerically diagonalizing the field-free relative-motion Hamiltonian on a two-dimensional square real-space grid. The kinetic-energy operator is approximated using second-order central finite differences, and the domain size is chosen large enough that the wave functions decay to negligible values at the boundary. Throughout, we use parameters representative of TMD electron-hole bilayers \cite{kamban_2020}, with effective masses chosen to be representative of a \eq{\mr{WSe_2/MoSe_2}} heterostructure \cite{yeh_2015, mostaani_2017, dai_2024, fu_2026}. The specific values used in each calculation are given in the corresponding figure captions.

\begin{figure}[htb]
    \includegraphics[width=0.95\linewidth]{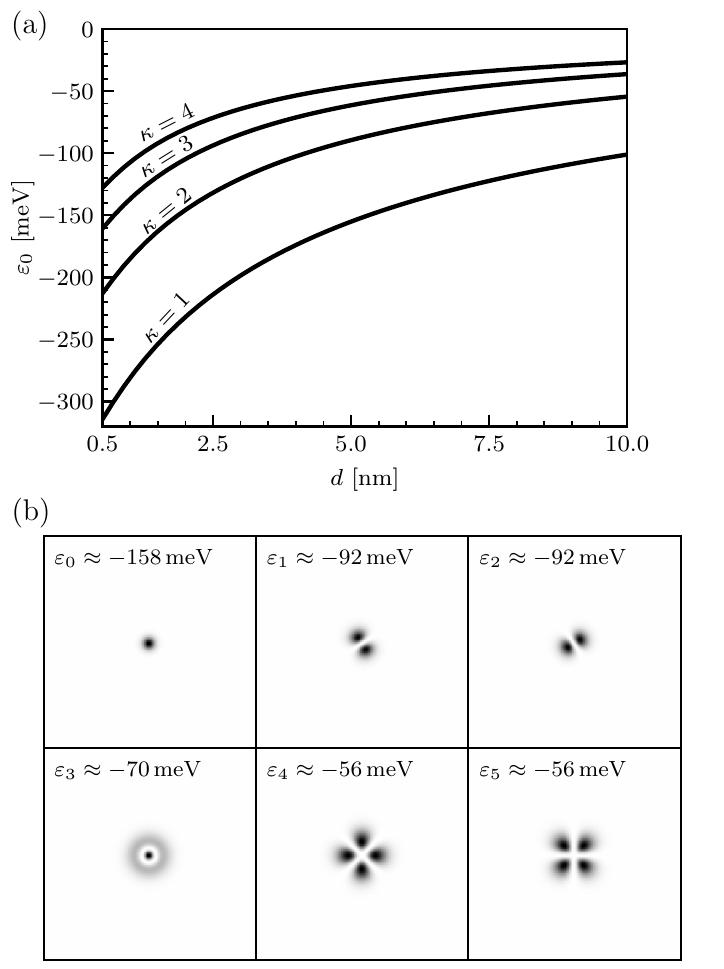}
    \caption{Interlayer-exciton properties at zero electric field: (a) ground-state energy as a function of layer spacing for several values of the effective dielectric constant, and (b) the lowest few eigenenergies along with the corresponding real-space probability densities of the relative coordinate. The parameters used are \eq{r_0 = \SI{8}{\nano\meter}}, \eq{m_\mr{e} = 0.8m_0}, and \eq{m_\mr{h} = 0.4m_0}. In panel (b), we further used \eq{d = \SI{1}{\nano\meter}} and \eq{\kappa = 2.5}.}
    \label{fig:exciton_properties}
\end{figure}

Figure~\ref{fig:exciton_properties}~(a) shows the numerically computed ground-state energy \eq{\varepsilon_0} as a function of interlayer spacing \eq{d} for several values of the effective dielectric constant \eq{\kappa}. The binding energy \eq{\varepsilon_\mr{b} = -\varepsilon_0} decreases with increasing \eq{d} because larger separations weaken the electron-hole attraction. Increasing \eq{\kappa} further reduces the binding energy due to stronger dielectric screening. Accordingly, the strongest binding occurs for small \eq{d} and small \eq{\kappa}. For the parameters used in Fig.~\ref{fig:exciton_properties}~(a), we find a binding energy reaching \eq{\varepsilon_b \approx \SI{300}{\milli\electronvolt}}, while for larger \eq{d} and \eq{\kappa} it drops to values comparable to the room temperature thermal energy \eq{k_\mr{B} T_\mr{room} \approx  \SI{26}{\milli\electronvolt}}. These values are consistent with those reported in Ref.~\cite{kamban_2020} and allow us to access both the strongly and weakly bound regimes.

Figure~\ref{fig:exciton_properties}~(b) shows the lowest few numerically computed eigenenergies together with the corresponding real-space probability densities. Since the field-free relative-motion Hamiltonian is rotationally symmetric, we label the eigenstates by a principal quantum number \eq{n} and an angular-momentum quantum number \eq{m}. The wave functions are of the form \eq{\psi_{n,m}(r, \varphi) = R_{n,m}(r) e^{i m \varphi}}, where \eq{(r, \varphi)} are polar coordinates of the relative coordinate. States with \eq{\pm m} are degenerate because the field-free relative-motion Hamiltonian only depends on \eq{\abs{m}}. The degenerate eigenfunctions shown in Fig.~\ref{fig:exciton_properties}~(b) correspond to real linear combinations of the \eq{\pm m} partners, which lead to lobe patterns rather than perfect azimuthal symmetry. The ground-state probability density is nodeless and isotropic, consistent with an \eq{s}-like exciton (\eq{m = 0}). The next two states are degenerate (\eq{\varepsilon_1 = \varepsilon_2}) and show a two-lobe pattern with a single nodal line, characteristic of \eq{p}-like excitons (\eq{m = \pm 1}). The probability density with energy \eq{\varepsilon_3} is isotropic with additional radial structure, consistent with an excited \eq{s}-like exciton. The next pair with energy \eq{\varepsilon_4 = \varepsilon_5} shows a four-lobe pattern with two nodal lines, characteristic of \eq{d}-like excitons (\eq{m = \pm 2}). These excited states provide the relevant intermediate levels for the IFE because the electric-dipole coupling connects states with different \eq{m}, with selection rule \eq{\Delta m = \pm 1}. Starting from the ground state, which carries no angular momentum \eq{(m = 0)}, we therefore expect virtual transitions to the \eq{p}-like \eq{(m = \pm 1}) states to dominate the IFE-induced angular momentum. Since the \eq{m=+1} and \eq{m=-1} channels carry opposite angular momenta, a finite IFE requires an imbalance between them. This imbalance is induced here by the helicity of the applied field. 

\subsection{Driven dynamics}
We now compute the time-dependent driven dynamics that generate the IFE-induced orbital magnetic moment. We start from the orbital magnetic moment of the electron-hole pair, \eq{M = \sum_{\alpha} q_\alpha (\bm{r}_\alpha \cross \bm{p}_\alpha)_z / (2m_\alpha)}, and rewrite it in relative and COM coordinates. Disregarding the COM contribution gives
\begin{equation} \label{eq:M_t}
    M = \mu_B \left(\frac{m_0}{m_\mr{h}} - \frac{m_0}{m_\mr{e}}\right) \frac{L}{\hbar} ,
\end{equation}
where \eq{L = (\bm{r} \cross \bm{p})_z} is the relative motion out-of-plane angular momentum operator and \eq{\mu_B = e \hbar / (2m_0)} is the Bohr magneton with the bare electron mass \eq{m_0}. Equation~\eqref{eq:M_t} shows that a nonzero orbital magnetization requires electron-hole effective-mass asymmetry as anticipated from the classical two-component Drude model. We obtain the driven dynamics by numerically solving the time-dependent Schr{\'o}dinger equation using Visscher's method \cite{visscher_1991}. The system is initialized in the ground state found above, and the rotating electric field is adiabatically turned on over three driving periods, followed by evolution for seven additional periods. Throughout, we compute the expectation value \eq{\expval{M}\!(t)} using Eq.~\eqref{eq:M_t}. After the field ramp, the system evolves in an approximately periodic driven state, and the IFE-induced magnetization corresponds to a cycle-averaged value in this state: \eq{\expval{M}_{nT}} with the average taken over \eq{n} periods \eq{T=2\pi/\omega} of the driving field. 

\begin{figure}[htb]
    \centering
    \includegraphics[width=0.95\linewidth]{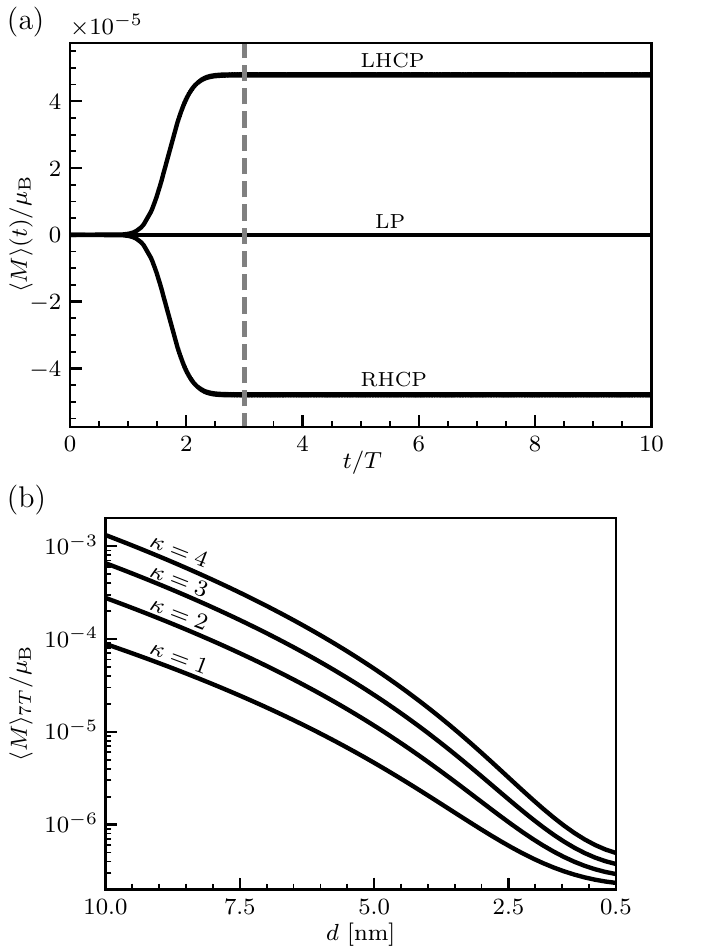}
    \caption{Orbital magnetization of the interlayer exciton under an oscillating electric field: (a) time dependence of the induced magnetization for linear polarization (LP), right-handed circular polarization (RHCP), and left-handed circular polarization (LHCP). The dashed line marks the end of the adiabatic ramp-up of the electric field. (b) Time-averaged magnetization in the periodically driven state as a function of layer spacing for several values of the effective dielectric constant. In both panels, we use an electric field with angular frequency \eq{\omega = \SI{e12}{\radian\per\second}} and amplitude \eq{E_0 = \SI{e5}{\volt\per\meter}}. The material parameters are \eq{r_0 = \SI{8}{\nano\meter}}, \eq{m_\mr{e} = 0.8m_0}, and \eq{m_\mr{h} = 0.4m_0}. In panel (a), we further use \eq{d = \SI{5}{\nano\meter}} and \eq{\kappa = 4}; panel (b) is computed with LHCP.}
    \label{fig:ife_quantum}
\end{figure}

Figure~\ref{fig:ife_quantum}~(a) shows the numerically computed time-dependent orbital magnetization during the adiabatic ramp-up of the rotating electric field and the subsequent periodically driven state after the dashed line. For circular polarization, the magnetization grows from zero during the ramp and saturates afterward, whereas for linear polarization it remains zero. This is consistent with the helicity dependence expected for the IFE. Figure~\ref{fig:ife_quantum}~(b) shows the dependence of the time-averaged magnetization on the interlayer spacing \eq{d} for several values of the effective dielectric constant \eq{\kappa}. The magnetization increases rapidly with increasing \eq{d}, consistent with a weakening of the electron-hole coupling as the layer separation increases. Increasing \eq{\kappa} enhances the time-averaged magnetization because the reduced electron-hole attraction makes the relative motion easier to drive.

\subsection{Time-dependent perturbation theory}
To gain analytical insight into the previous numerical results, we use time-dependent perturbation theory to derive an approximate expression for the time-averaged magnetization. We expand the time-dependent state in the eigenbasis of the field-free relative-motion Hamiltonian as
\begin{equation}
    \ket{\Psi(t)} = \sum_{n, m} c_{n, m}(t) \ket{n, m} e^{-i\varepsilon_{n, m} t/ \hbar},
\end{equation}
where \eq{\abs{c_{n,m}(t)}^2} gives the occupation probability of state \eq{\ket{n,m}}. The relative-motion orbital angular momentum is diagonal in this basis, and its expectation value reads
\begin{equation} \label{eq:L_t}
	\expval{L}\!(t) =\sum_n \sum_{m > 0}  \hbar m \left[ \abs{c_{n, m}(t)}^2 - \abs{c_{n, -m}(t)}^2 \right].
\end{equation}
This shows that a nonzero orbital angular momentum requires an imbalance between the \eq{+m} and \eq{-m} sectors. Here, it is the helicity of the applied electric field that generates this imbalance. We obtain the occupation probabilities perturbatively by expanding in powers of the dipole coupling, \eq{c_{n,m} = \sum_{i=0}^\infty c_{n, m}^{(i)}(t)}. Retaining terms up to second order in the electric field gives
\begin{equation}
\begin{split}
	\abs\big{c_{n,m}(t)}^2 &= \abs\big{c_{n,m}^{(0)}}^2 + \abs\big{c^{(1)}_{n, m}(t)}^2
	\\
	&+ 2 \Re \big[c_{n, m}^{(0)} c^{(1)*}_{n, m}(t) + c^{(0)}_{n, m} c^{(2)*}_{n, m}(t) \big].
\end{split}
\end{equation}
We initialize the system in the field-free ground state: \eq{c^{(0)}_{n,m} = \delta_{n, 0} \delta_{m, 0}}. Consequently, the terms involving \eq{c^{(0)}_{n, m}} do not contribute to Eq.~\eqref{eq:L_t}. The leading contribution thus comes from \eq{|c^{(1)}_{n, \pm m}|^2}. Using time-dependent perturbation theory as detailed in Appendix~\ref{app:time-dependent}, we obtain
\begin{equation} \label{eq:perturbation_theory}
	\abs\big{c_{n,m}^{(1)}}^2 = \frac{e^2}{4\hbar^2} \abs\Big{\sum_\sigma \mel{n,m}{r_\sigma}{0,0} f_\sigma(t)}^2
\end{equation}
where \eq{\sigma \in \{+, -\}}, the circular components of the position operator are \eq{r_\pm = (r_x \pm i r_y) / \sqrt{2}}, and the time dependence is governed by
\begin{equation}
	f_\mp(t) = \frac{E^\omega_\pm e^{i \omega t}}{\Omega_{n,m} + \omega} + \frac{E^{\omega *}_\mp e^{-i \omega t}}{\Omega_{n, m} - \omega}.
\end{equation}
The angular frequency of the drive is \eq{\omega} and the transition energy is defined with respect to the ground state as \eq{\hbar \Omega_{n, m} = \varepsilon_{n, m} - \varepsilon_{0, 0}}. Importantly, the circular components of the position operators carry angular momentum and impose selection rules: \eq{r_\pm} connects only states with \eq{m \to m\pm1} within the dipole approximation. Moreover, the field-free relative-motion Hamiltonian is rotationally invariant with a real central potential, which implies \eq{\abs{\!\mel{n,1}{r_+}{0,0}}^2 = \abs{\!\mel{n,-1}{r_-}{0,0}}^2}. Using these relations, we insert Eq.~\eqref{eq:perturbation_theory} into Eq.~\eqref{eq:L_t} and obtain
\begin{equation} \label{eq:result_L}
	\expval{L}\!(t) = \frac{2e^2}{\hbar} \sum_n \abs{\!\mel{n, 1}{r_+}{0,0}}^2 \frac{\omega \Omega_{n,1}}{\left(\Omega_{n, 1}^2 - \omega^2 \right)^2} E^2_0 \mathcal{P},
\end{equation}
where the helicity \eq{\mathcal{P}} and amplitude \eq{E_0} are defined in and below Eq.~\eqref{eq:helicity}. The oscillatory terms at \eq{\omega} and \eq{2\omega} cancel and do not contribute at this order. The leading response is thus time independent and we identify it with the IFE-induced magnetization because of the factor \eq{E_0^2 \mathcal{P}} analogous to Eq.~\eqref{eq:ife_depend}. This result is also consistent with the numerical results in Fig.~\ref{fig:ife_quantum}~(a), where the magnetization in the periodically driven state is essentially time independent, with only small rapidly oscillating corrections arising from higher-order contributions.

We now consider off-resonant driving where \eq{\omega \ll \Omega_{n, 1}} and expand the frequency-dependent factor in Eq.~\eqref{eq:result_L} to leading order: \eq{\omega \Omega_{n, 1} / (\Omega_{n, 1}^2 - \omega^2)^2 \approx \omega / \Omega^3_{n, 1}}. For this purpose, it is convenient to introduce an effective transition frequency
\begin{equation} \label{eq:effective_gap}
	\Omega_\mr{eff}^{-3} = \frac{\sum_n \abs{\mel{n, 1}{r_+}{0,0}}^2 \Omega^{-3}_{n, 1}}{ \sum_n \abs{\mel{n, 1}{r_+}{0,0}}^2},
\end{equation}
which collects the transition frequency dependence into a single scale while retaining the contributions from all coupled excited states. We note that a simple estimate for the effective transition frequency is the transition frequency to the first excited state. The denominator in Eq.~\eqref{eq:effective_gap} can be evaluated exactly using the selection rules together with basis completeness,
\begin{equation}
	\sum_n \abs{\! \mel{n,1}{r_+}{0,0}}^2 = \mel{0,0}{r_- r_+}{0,0} = \frac{1}{2} r^2_\mr{rms},
\end{equation}
where \eq{r^2_\mr{rms} = \mel{0,0}{r^2}{0,0}}. Defining the effective transition energy as \eq{\Delta_\mr{eff} = \hbar \Omega_\mr{eff}}, and expressing the field amplitude in terms of a Rabi frequency scale \eq{\hbar \Omega_\mr{R} = e r_\mr{rms} E_0}, we insert the off-resonant form of Eq.~\eqref{eq:result_L} into Eq.~\eqref{eq:M_t} and obtain
\begin{equation} \label{eq:approximate}
	\expval{M}\!(t) \approx \mu_\mr{B} \frac{\hbar \omega}{\Delta_\mr{eff}} \left(\frac{\hbar \Omega_\mr{R}}{\Delta_\mr{eff}} \right)^2 \left( \frac{m_0}{m_\mr{h}} - \frac{m_0}{m_\mr{e}}  \right) \mathcal{P}.
\end{equation}
Equation~\eqref{eq:approximate} shows how the response depends on the relevant energy scales. The relevant energy scales are determined by the ratios of the photon energy and the Rabi energy to the effective transition energy. The IFE response is thus enhanced for more weakly bound excitons, consistent with the numerical results in Fig.~\ref{fig:ife_quantum}~(b). We understand this intuitively as more weakly bound excitons are more spatially extended, allowing the rotating field to induce larger relative orbital motion. This is analogous to recent work on orbital IFE in Rydberg systems, where spatially extended bound-state wave functions enhance the induced orbital magnetization \cite{wong_2025}. Finally, the helicity and effective-mass asymmetry required for a nonzero IFE are consistent with the classical model. We evaluate Eq.~\eqref{eq:approximate} using representative interlayer exciton parameters: \eq{r_\mr{rms} \approx \SI{2}{\nano\meter}}, \eq{\Delta_\mr{eff} \approx \SI{40}{\milli\electronvolt}}, \eq{m_\mr{e} = 0.8 m_0}, \eq{m_\mr{h} = 0.4 m_0}, and LHCP (\eq{\mathcal{P} = 1}) terahertz driving field with energy \eq{\hbar \omega \approx \SI{0.66}{\milli\electronvolt}}. This gives \eq{M \approx \mu_\mr{B} [E_0 / (\SI{e8}{\volt\per\meter})]^2}, which for the field strength used in the numerical simulations \eq{E_0 = \SI{e5}{\volt\per\meter}} matches the simulated order of magnitude at small interlayer separations.

\section{Summary} \label{sec:summary}
We have shown that circularly polarized light can generate a rectified orbital magnetization in electron-hole bilayers, thereby demonstrating the concept of quantum printing as applied to neutral electron fluids. The effect requires an asymmetry between the carriers in the two layers. We consider the specific case of transition-metal dichalcogenide bilayers as the reference platform because of their tunability, where electron–hole effective-mass asymmetry provides the most relevant mechanism. The induced magnetization can therefore remain nonzero even in fully compensated electron-hole bilayers.

In the weak-coupling regime, a two-component Drude model with interlayer Coulomb drag predicts a sizable IFE for charge-balanced bilayers with representative effective-mass ratios. The response is suppressed as the interlayer coupling grows and the electron and hole motions become locked. In the strong-coupling regime, described by a hydrogenic model of the interlayer exciton with the Rytova-Keldysh interaction, numerical time evolution yields a smaller but finite rectified magnetization. Time-dependent perturbation theory identifies the mechanism: the helicity of the drive biases virtual transitions toward exciton states carrying orbital angular momentum. On this basis, we obtained a compact off-resonant expression for the IFE in the excitonic regime.

Because the rectified magnetization is strongly sensitive to both the electron-hole asymmetry and the interlayer coupling strength, the IFE can offer a useful probe of interlayer physics in these systems. Helicity-resolved measurements could track the crossover from weakly coupled carrier motion to interlayer exciton dynamics. A natural experiment would combine terahertz pumping with a magneto-optical probe \cite{conte_2015, hsu_2015, yoon_2022, policht_2023}. Our analysis rests on simplified effective-mass descriptions, while quantitative material-specific predictions would require a microscopic treatment of electronic structure, screening, scattering, and excitonic relaxation dynamics.

\begin{acknowledgments}
We are grateful to C. Ahn, G. Aeppli, R. Castillo-Garza, H.T. Chen, M. Jain, S.-Z. Lin, I. Sochnikov, and J.-X. Zhu for useful conversations. E.S. thanks the Nordic Institute for Theoretical Physics for hospitality and Olav Syljuåsen for useful discussions on the numerical simulations.

E.S. acknowledges support from the Research Council of Norway through Grant No. 262633 ``Center of Excellence on Quantum Spintronics''. P.W. and A.B. were supported by the U.S. Department of Energy, Office of Science, Office of Basic Energy Sciences, under award number DE-SC-0025580. G.C. and P.W. were also supported by the European Research Council under the European Union Seventh Framework ERS-2018-SYG 810451 HERO. E.I.A. acknowledges support from the International Max Planck Research School and thanks the Nordic Institute for Theoretical Physics for hospitality. A.R. acknowledges support from the European Research Council (ERC-2024-SyG-101167294; UnMySt), the Cluster of Excellence “Advanced Imaging of Matter” (AIM), Grupos Consolidados (IT1453-22), and the Max Planck–New York City Center for Non-Equilibrium Quantum Phenomena. The Flatiron Institute is a division of the Simons Foundation. 
\end{acknowledgments}

\section*{Data Availability}
The source code and numerical data supporting the findings of this study are publicly available \cite{syljuaasen_2026}.

\appendix 
\section{Hydrodynamic IFE} \label{appendix:ife}
The strategy for obtaining the IFE-induced magnetization within a hydrodynamic description is to identify the rectified solenoidal component of the charge current, from which the corresponding magnetization can be extracted  \cite{karpman_1982, hertel_2006}. This is conveniently done by introducing a weak spatial dependence in the applied electric field. However, after isolating the rectified magnetization, we take the uniform limit to obtain the magnetization that remains for a homogeneous drive. We now rederive this procedure following Refs.~\cite{karpman_1982, hertel_2006}, in a form suitable for the bilayer system considered.

We expect the response to be periodic at the same frequency \eq{\omega} as the applied electric field, and therefore expand the density and velocity in harmonics of \eq{\omega}. To first order,
\begin{align} 
	n_\alpha(\bm{r}, t) = n_\alpha^0(\bm{r}) + \Re\left[n_\alpha^\omega(\bm{r}) e^{i\omega t}\right], \label{eq:density_fourier}
	\\
	\bm{v}_\alpha(\bm{r}, t) = \bm{v}_\alpha^0(\bm{r}) + \Re\left[\bm{v}_\alpha^\omega(\bm{r}) e^{i\omega t}\right], \label{eq:velocity_fourier}
\end{align}
with \eq{n^{-\omega}_\alpha = n^{\omega *}_\alpha} and \eq{\bm{v}^{-\omega}_\alpha = \bm{v}^{\omega *}_\alpha} because both fields are real-valued in the time domain. We consider an electric field that is purely oscillatory in time, and in the uniform-field limit we therefore set the time-averaged drift to zero, \eq{\bm{v}^0_\alpha = 0}, and take the equilibrium density to be uniform, \eq{n^0_\alpha(\bm{r}) = n^0_\alpha}. Substituting Eqs.~(\ref{eq:density_fourier}, \ref{eq:velocity_fourier}) into the charge current \eq{\bm{J}_\alpha(\bm{r}, t) = q_\alpha n_\alpha(\bm{r}, t) \bm{v}_\alpha(\bm{r}, t)} yields a rectified contribution
\begin{equation} \label{eq:rectified_current}
    \bm{J}^{0}_\alpha(\bm{r}) = \frac{q_\alpha}{4} \left[n^{\omega}_\alpha(\bm{r}) \bm{v}^{\omega *}_\alpha(\bm{r}) + n^{\omega *}_\alpha(\bm{r}) \bm{v}^{\omega}_\alpha(\bm{r}) \right].
\end{equation}
Higher harmonics would also contribute to the time-averaged response in a fully nonlinear treatment. However, since the velocity response from the equations of motion is linear in the electric field, we keep only this leading rectified term. The first-harmonic density follows from the continuity equation in Eq.~\eqref{eq:continuity},
\begin{equation} \label{eq:intermediate}
    n^{\omega}_\alpha(\bm{r}) = \frac{i}{\omega} n^{0}_\alpha \div \bm{v}^{\omega}_\alpha(\bm{r}).
\end{equation}
Inserting Eq.~\eqref{eq:intermediate} into Eq.~\eqref{eq:rectified_current} gives
\begin{equation}
	\bm{J}^0_\alpha = \frac{iq_\alpha n^0_\alpha}{4\omega} \left\{ \left[\div \bm{v}^\omega_\alpha(\bm{r}) \right] \bm{v}^{\omega *}_\alpha(\bm{r}) - \left[\div \bm{v}^{\omega *}_\alpha(\bm{r}) \right] \bm{v}^\omega_\alpha(\bm{r}) \right\}.
\end{equation}
Then, using the vector identity 
\begin{equation}
	(\div \bm{a}) \bm{b} - (\div \bm{b}) \bm{a}  = - \curl (\bm{a} \cross \bm{b}) - (\bm{a} \vdot \bm{\nabla}) \bm{b} + (\bm{b} \vdot \bm{\nabla}) \bm{a}
\end{equation}
valid for arbitrary vectors \eq{\bm{a}} and \eq{\bm{b}}, the rectified current can be decomposed as 
\begin{equation}
	\bm{J}^0_\alpha(\bm{r}) = \bm{\Gamma}^0_\alpha(\bm{r}) + \curl \bm{M}^0_\alpha(\bm{r}),
\end{equation}
where
\begin{align}
	\bm{\Gamma}^0_\alpha(\bm{r}) &= \frac{q_\alpha n^0_\alpha}{2\omega} \Im\left\{ \left[ \bm{v}^\omega_\alpha(\bm{r}) \vdot \bm{\nabla} \right] \bm{v}^{\omega *}_\alpha(\bm{r}) \right\},
    \\
 	\bm{M}^0_\alpha(\bm{r}) &= \frac{q_\alpha n_\alpha^0}{4\omega} \Im\left[\bm{v}^\omega_{\alpha}(\bm{r}) \cross \bm{v}^{\omega *}_{\alpha}(\bm{r}) \right].  \label{eq:rectified_magnetization}
\end{align} 
The term \eq{\bm{\Gamma}^0_\alpha} describes a rectified current arising from spatial gradients of the driving field \cite{karpman_1982, hertel_2006}. These gradients generate ponderomotive forces, which can drive circulating currents and associated orbital magnetization, but vanish in the uniform-field limit \cite{karpman_1982, hertel_2006}. The term \eq{\bm{M}^0_\alpha} represents the IFE-induced magnetization associated with a homogeneous applied electric field. 

We note that the rectified magnetization in Eq.~\eqref{eq:rectified_magnetization} can also be derived by considering the orbital magnetic moment for a single point charge,
\begin{equation} \label{eq:magnetic_moment}
    \bm{m}_\alpha(t) = \frac{q_\alpha}{2} \bm{r}_\alpha(t) \cross \bm{v}_\alpha(t),
\end{equation}
where \eq{\bm{r}_\alpha} is the particle displacement. Expanding the displacement to first order as \eq{\bm{r}_\alpha = \bm{r}_\alpha^0 + \Re(\bm{r}^\omega_\alpha e^{i\omega t})}, and using \eq{\partial_t \bm{r}_\alpha(t) = \bm{v}_\alpha(t)}, gives
\begin{equation}
\begin{split}
    \bm{m}_\alpha(t) = \frac{q_\alpha}{2} &\left(\bm{r}_\alpha^0 + \frac{1}{2i\omega} \bm{v}^\omega_\alpha e^{i\omega t} - \frac{1}{2i\omega} \bm{v}^{\omega *}_\alpha e^{-i\omega t} \right)
    \\
    \cross  &\left( \frac{1}{2} \bm{v}^\omega_\alpha e^{i\omega t} + \frac{1}{2} \bm{v}^{\omega *}_\alpha e^{-i\omega t} \right).
\end{split}
\end{equation}
The rectified part can now be identified, and we denote it by \eq{\bm{m}^0_\alpha}. Multiplying by the time-averaged density gives \eq{n^0_\alpha \bm{m}^0_\alpha = \bm{M}^0_\alpha}, which reproduces Eq.~\eqref{eq:rectified_magnetization}. We therefore interpret the IFE-induced magnetization in the hydrodynamic description as the sum of the individual time-averaged orbital magnetic moments of the carriers.

\section{Classical dipole limit} \label{app:classical_dipole_limit}
To gain intuition for the dipolar limit, we consider the classical dynamics of a rigid dipole driven by circularly polarized light. We work in center-of-mass and relative coordinates and replace the interacting electron-hole Hamiltonian in Eq.~\eqref{eq:relative_H} by the classical Hamiltonian
\begin{equation}
    H(t) = \frac{\bm{P}^2}{2(m_\mr{e} + m_\mr{h})}  + \frac{\bm{p}^2}{2 \mu} + e \bm{r} \vdot \bm{E}(t).
\end{equation}
For simplicity, we use the rotating electric field \eq{\bm{E}(t) = E_0 [\cos(\omega t) \hat{\bm{x}} + \sin(\omega t) \hat{\bm{y}}]} and constrain the relative coordinate to
\begin{equation} \label{eq:r_constrained}
    \bm{r} = d_\perp \left[\cos(\theta) \hat{\bm{x}} + \sin(\theta) \hat{\bm{y}}\right] + d_z \hat{\bm{z}},
\end{equation}
where \eq{d_\perp} is the fixed in-plane dipole separation, \eq{d_z} is the fixed interlayer distance, and \eq{\theta} is the in-plane angle of the dipole. The relative-motion Hamiltonian is then
\begin{equation}
    H_\mr{rel}(t) = \frac{p_\theta^2}{2 \mu d_\perp^2} + e d_\perp E_0 \cos(\theta - \omega t),
\end{equation}
where \eq{p_\theta} is the momentum conjugate to \eq{\theta}. The corresponding equations of motion are
\begin{align}
   \dot{\theta} &= \frac{p_\theta}{\mu d_\perp^2},
   &
   \dot{p_\theta} &= e d_\perp E_0 \sin(\theta - \omega t),
\end{align}
which combine to give
\begin{equation}
    \ddot{\theta} - \frac{e E_0}{\mu d_\perp} \sin(\theta - \omega t) = 0.
\end{equation}
We obtain the pendulum equation by introducing the phase difference \eq{\phi = \theta - \omega t + \pi}, which gives
\begin{equation}
    \ddot{\phi} + \Omega^2 \sin(\phi) = 0, \qq{where} \Omega^2 = \frac{e E_0}{\mu d_\perp}.
\end{equation}
The magnetic moment arising from the relative motion follows from Eq.~\eqref{eq:M_t}:
\begin{equation}
    M_z = \frac{e}{2} \left(\frac{m_\mr{e} - m_\mr{h}}{m_\mr{e} + m_\mr{h}}\right) (\bm{r} \cross \dot{\bm{r}})_z,
\end{equation}
where inserting Eq.~\eqref{eq:r_constrained} yields \eq{(\bm{r} \cross \dot{\bm{r}})_z = d_\perp^2 (\dot{\phi} + \omega)}. We are interested in the time-averaged magnetization over \eq{n} periods \eq{T}, where \eq{\langle \dot{\phi} \rangle_{nT} = [\phi(nT) - \phi(0)] / (nT)} which tends to zero for bounded motion, such that
\begin{equation}
    \expval{M_z}_T = \frac{e}{2} \left(\frac{m_\mr{e} - m_\mr{h}}{m_\mr{e} + m_\mr{h}} \right) d_\perp^2 \omega.
\end{equation}
This result shows that the induced static magnetic moment requires both a finite in-plane separation between the charges forming the dipole and electron-hole mass asymmetry.

\section{Time-dependent perturbation theory} \label{app:time-dependent}
We use standard time-dependent perturbation theory \cite{hemmer_2005} to derive the first-order correction to the wave function induced by a rotating monochromatic electric field. We write the Hamiltonian as the sum of an equilibrium part and a time-dependent perturbation, \eq{H(t) = H_0 + H'(t)}, and expand the state as  
\begin{equation}
	\ket{\Psi(t)} = \sum_{\mu} c_{\mu}(t) \ket{\mu} e^{-i \varepsilon_{\mu} t / \hbar}.
\end{equation}
Here, \eq{\ket{\mu}} denotes an eigenstate of \eq{H_0} with eigenvalue \eq{\varepsilon_\mu}, and \eq{c_\mu(t)} is the corresponding time-dependent amplitude. The time evolution follows from the Schr{\"o}dinger equation,
\begin{equation}
	\dot{c}_\mu(t) = \frac{1}{i \hbar} \sum_{\nu} \mel{\mu}{H'(t)}{\nu} e^{i \omega_{\mu \nu} t} c_{\nu}(t),
\end{equation}
where \eq{\hbar \omega_{\mu \nu} = \varepsilon_\mu - \varepsilon_\nu}. The above equation can be solved perturbatively in powers of \eq{H'(t)} by expanding \eq{c_\mu(t) = \sum_{i = 0}^\infty c_\mu^{(i)}(t)}, where the first two terms are
\begin{align}
	c_\mu^{(0)} &= \lim_{t \to -\infty} c_\mu(t),
	\\
	c_\mu^{(1)}(t) &= \frac{1}{i\hbar} \sum_{\nu} c_{\nu}^{(0)} \! \int\limits_{-\infty}^t \!\!\! \dd t' \mel{\mu}{H'(t')}{\nu} e^{i \omega_{\mu \nu} t'} . \label{eq:first_order_correction}
\end{align}
The perturbation is the electric-dipole interaction, 
\begin{equation}
	H'(t) = e \bm{r} \vdot \bm{E}(t) e^{\eta t},
\end{equation}
where the limit \eq{\eta \to 0^+} implements an adiabatic switch-on of the electric field. As described in the main text, we use a circular basis and decompose the interaction as \eq{\bm{r} \vdot \bm{E}(t) = r_+ E_-(t) + r_- E_+(t)}, with \eq{r_\pm = (x \pm i y) / \sqrt{2}} and \eq{E_\pm(t) = [E_x(t) \pm i E_y(t)] / \sqrt{2}}. With our convention \eq{\bm{E}(t) = \Re(\bm{E}^\omega e^{i\omega t})}, this gives
\begin{equation}
\begin{split}
	E_\pm(t) &= \left[\Re(E^\omega_x e^{i\omega t}) \pm i \Re(E^\omega_y e^{i\omega t}) \right] / \sqrt{2}
	\\ 
	&= \left(E^\omega_\pm e^{i\omega t} + E^{\omega *}_{\mp} e^{-i\omega t} \right) / 2,
\end{split}
\end{equation} \\
where \eq{E_\pm^\omega = (E^\omega_x \pm i E^\omega_y) / \sqrt{2}}. Inserting the perturbation into Eq.~\eqref{eq:first_order_correction} and carrying out the temporal integral finally yields
\begin{equation}
\begin{split}
	c_\mu^{(1)}(t) = -\frac{e}{2\hbar} &\sum_\nu c_\nu^{(0)} e^{i \omega_{\mu \nu} t} e^{\eta t} 
	\\
	\times \bigg[ \mel{\mu}{r_+}{\nu} &\left( \frac{E^\omega_- e^{i \omega t}}{\omega_{\mu \nu} + \omega - i \eta}  + \frac{E^{\omega *}_+ e^{-i \omega t}}{\omega_{\mu \nu} - \omega - i\eta} \right)
	\\
	+\mel{\mu}{r_-}{\nu} &\left( \frac{E^\omega_+ e^{i \omega t}}{\omega_{\mu \nu} + \omega - i\eta}  + \frac{E^{\omega *}_- e^{-i \omega t}}{\omega_{\mu \nu} - \omega - i \eta} \right) \bigg].
\end{split}
\end{equation}
The denominators remain finite when \eq{\eta \to 0^+} since we focus on the off-resonant oscillatory response.

\bibliography{references}
\end{document}